\documentclass{article}
\usepackage[utf8]{inputenc}

\title{Collective computation in a network with distributed information}

\author{A. C\'ordoba\\Departamento de F\'isica de la Materia Condensada, Universidad de Sevilla,  \\ 
        P. O. Box 1065, 41080 Sevilla, Spain\\cordoba{@}us.es\\
        D. Aguilar-Hidalgo\\Max Planck Institute for the Physics of Complex Systems, \\
         N\"{o}thnitzer Str. 38, 01187 - Dresden, Germany\\
         M. C. Lemos\\Departamento de F\'isica de la Materia Condensada, Universidad de Sevilla,  \\ 
        P. O. Box 1065, 41080 Sevilla, Spain}

\date{}

\usepackage{amssymb}
\usepackage{amsmath}
\usepackage{longtable}
\usepackage{geometry}
\usepackage{pifont}
\usepackage{graphicx}
\usepackage{setspace}
\usepackage{tabulary}

\begin{document}
\maketitle

\begin{abstract}
We analyze a distributed information network in which each node has access to the information contained in a limited set of nodes (its neighborhood) at a given time. A collective computation is carried out in which each node calculates a value that implies all information contained in the network (in our case, the average value of a variable that can take different values in each network node). The neighborhoods can change dynamically by exchanging neighbors with other nodes. The results of this collective calculation show rapid convergence and good scalability with the network size. These results are compared with those of a fixed network arranged as a square lattice, in which the number of rounds to achieve a given accuracy is very high when the size of the network increases. The results for the evolving networks are interpreted in light of the properties of complex networks and are directly relevant to the diameter and characteristic path length of the networks, which seem to express "small world" properties.
\end{abstract}




\section{Introduction}

We propose a model for a distributed information network in which each node at a given time can obtain information from a limited number of other nodes to which it is connected. From the information residing in these nodes, each one of them can perform specific tasks (calculation, classification, etc.) involving all the information contained in the network nodes. In a system where information is distributed at different sites, the access to this information can be an expensive procedure if the number of sites is high. Moreover, if a set of $N$ nodes want to do at any given time a calculation of a magnitude involving information contained in all other ones, the direct access of all to all nodes requires a number of requests of order $N^2$, which is very high if $N$ is large.

Here we pose the problem of a set of $N$ elements (nodes), each characterized by the value of a magnitude $s$, in general different for each node, so that each and every one of the elements wish to calculate the average value of $s$ (or a function of $s$) in the set, by accessing at one time to information contained in other $q$ elements of the set ($q << N$) with which it is connected. This is in line with systems using epidemic protocols \cite{Eugster:2004,De:2009,Anagnostopoulos:2011} or gossip ones \cite{Kermarrec:2007,Islam:2009,Tang:2011}, such as the newscast protocol \cite{Voulgaris:2004}. In these systems the goal is not to enable both point-to-point communication between nodes, but rapid and efficient dissemination of information. The system intends to perform a specific collective task (for example, calculating the average value of a variable in the set of nodes, setting the position of each node in a ranking according to the set value of the variable, etc.) so that, eventually, all nodes have access to the result obtained from the set.  To do this we consider a network in which each element is connected to other $q$ ones (neighbors), from which it extracts information. The neighborhood of each element can be changed along the process of calculation, so that each node can exchange a neighbor randomly with another node. We discuss issues such as scalability and convergence, considering different sizes and data sets. We also compare the results obtained for the dynamically changing network with those obtained from the fixed static network. To do this we compare network configurations from different times of evolution with those corresponding to the initial fixed network (which may take the form of a conventional cellular automaton). 

\section{The model}

\begin{figure}
\begin{center}
  \includegraphics[width=\textwidth]{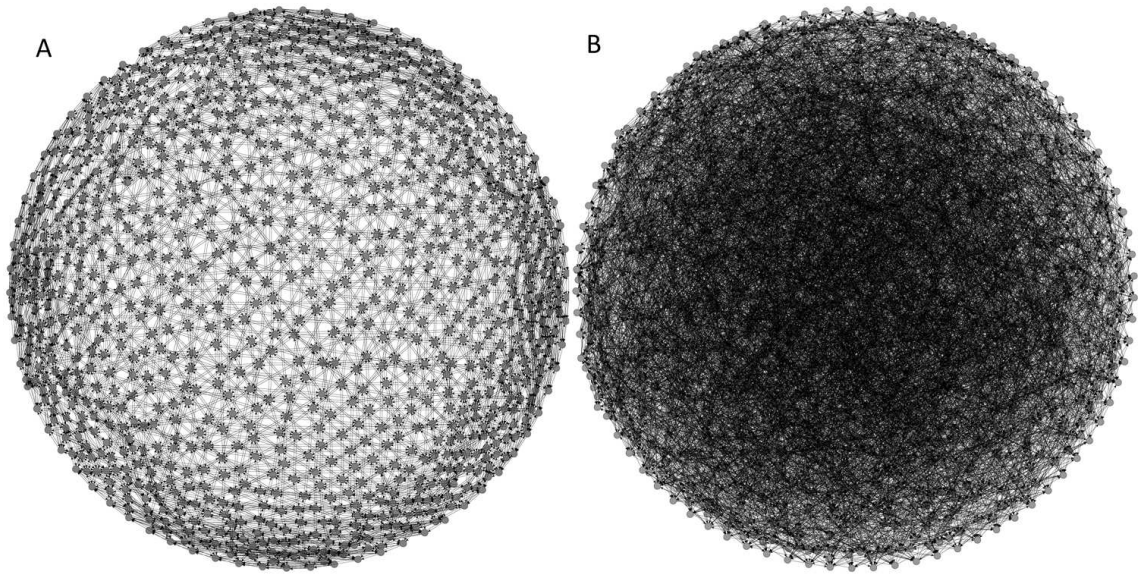}
  \end{center}
  \caption{Representation of 32x32 networks (1024 nodes and 8192 links). A) Network generated by the Cellular Automaton with Moore's neighborhood and used as an initial condition for the Evolutionary method. B) Evolutionary network representation. This last network is far less ordered than the cellular automaton case.}
  \label{fig:networks}
\end{figure}

In a gossip framework there is an exchange of information among system elements In this exchange of information, which is a dynamic process, one receives information from another (also can be a reciprocal exchange). In turn, the receiver of information can give information to other peers. Overall the transmission process basically comprises three aspects: peer selection, data exchanged or transmitted between peers and data processing. In our model we consider a network arranged in a two-dimensional lattice (we do this initially to compare with a conventional cellular automaton with Moore's neighborhood \cite{Wolfram:2002}) with $m \times n$ sites forming a square lattice. To each node is randomly assigned the value of a variable $s$, in general different for each node. The objective is that each node "knows" the average value of this variable in the whole at the end of the process. Each node has access to limited information on each step of calculation. To obtain specific results, it has been considered that each node, at each instant, can only store eight values of other nodes, and that initially each node has access to the data of the eight nearest neighbors in the lattice. This can be matched with a two-dimensional cellular automaton with Moore's neighborhood. Throughout the process the neighborhood of each site is not fixed, but each node exchanges a neighbor with another randomly chosen node. This implies a double selection (with appropriate calls to random numbers routines), one of them for the site of exchanging and the other for choosing the neighbor exchanged. The average value of the variable of each site and its eight neighbors (inputs) is assigned to the variable of the corresponding site for each system update. The updating of all sites is simultaneous, although it could also be carried out sequentially. Since in the course of evolution neighborhoods randomly change, the initial regular structure of the network is irrelevant (only holds the fact that each node has eight input and eight outputs, generally different). However, in order to formally create a by analogy with the above mentioned cellular automaton, we initially identify each node as a point on a square lattice $m\times n$ (Figure  \ref{fig:networks}A), but now, after the dynamic exchange of information, the system is turned into a complex directed network and connections do not generate a square lattice (Figure  \ref{fig:networks}B)\footnote{Figures  \ref{fig:networks} A and B are represented in a circular layout in order to compact the graph, as an orthogonal layout (square lattice like) is not friendly looking due to the high number of nodes.}.

To test the scalability we have considered various system sizes ($32 \times 32$, $100 \times 100$, $320 \times 320$ and $1000 \times 1000$). Also we have considered two data distributions: (a) a uniform random distribution, (b) a distribution of values grouped into four quadrants as shown schematically in Figure \ref{fig:grouped_values}. For comparison, as mentioned above, three cases are considered: the network that dynamically changes, the fixed network with the configuration that is reached at a given moment of the dynamical evolution and the fixed network with the initial square lattice structure (cellular automaton). In the latter two cases, the neighborhoods do not change and each node has permanent access to its pre-established eight neighbors.

We defined a parameter, $b$, to evaluate how the computation progresses in each system update, i.e., convergence to the desired value with a given accuracy.

\begin{equation}
b = \mbox{standard deviation} / \mbox{mean value} 
\end{equation}

We have imposed the condition that the calculation is stopped when the variation of $b$ between two successive updates is less than a preset value. We have also established a limited number of system updates. Moreover, using the Cytoscape software \cite{Shannon:2003,Assenov:2008}, we have analyzed some of the properties of the network [11-12] to establish the relationship between its topology and the degree of computation efficiency.

\section{Results and Discussion}

\begin{figure}
\begin{center}
  \includegraphics[width=10cm]{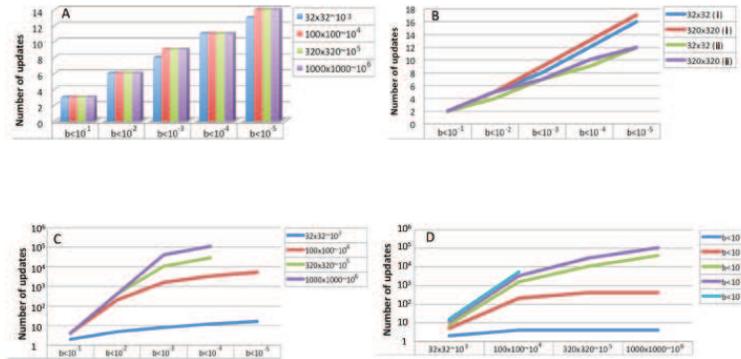}
  \end{center}
  \caption{(A) Number of updates required to reach the $b$ value in an evolutionary network for a randomly distributed data set. The bar graph shows a linear increase of the number of updates with the required precision and remains constant when increasing the network size. (B) Number of updates required to reach the $b$ value in two fixed networks for a randomly distributed data set: (i) Evolutionary network configuration when $b<10^{-1}$. (ii) Evolutionary network configuration when $b<10^{-5}$. As in the case of using evolutionary network, the number of updates needed to reach a certain precision in the calculus increases linearly with this precision. In this case, the increase in the network size makes the number of updates to remain nearly constant. (C, D) Number of updates required to reach the $b$ value applying a cellular automaton with Moore neighborhood for a randomly distributed data set. (C) varying $b$. (D) Varying the network size. Contrary to the two anterior cases of study, the use of cellular automata makes the number updates increase in an exponential tendency as the precision and the network size grow.}
  \label{fig:results1}
\end{figure}

We have taken the value of $b = 10^{-5}$ as the its limit of $b$ and have scored the number of updates required to reach a value of $b$ less than the successive powers of 10 from $10^{-1}$ to $10^{-5}$. The Table \ref{tab:tab1} shows the results for the evolving network for a set of data randomly distributed according to a uniform distribution. As can be seen, the number of updates is low and the scalability is very good, because for large variations in the size of the system the number of updates nearly remains the same. Each node has to perform a small number of updates. Since for each update of the entire system one must perform $N = m \times n$ updates of the values of the nodes, then, the total number of individual operations is virtually proportional to system size, i.e. escalates as $N$ (a direct calculation escalates as $N^{2}$ ). Figure \ref{fig:results1}A graphically show this behavior where the number of updates slowly increases when $b$ is diminished. It must be noticed that the number of updates remains the same for a certain $b$ independently of the network size.

\begin{table}
\begin{tabular}{|c|c|c|c|c|c|}
\hline
\multicolumn{6}{|c|}{Evolutionary Network}\\
\hline
$m \times n$ & $b<10^{-1}$ & $b<10^{-2}$ & $b<10^{-3}$ & $b<10^{-4}$ & $b<10^{-5}$ \\
\hline
$32 \times 32 \sim 10^3$ & 3 & 6 & 8 & 11 & 13\\
\hline
$100 \times 100 \sim10^4$ & 3 & 6 & 9 & 11 & 14\\
\hline
$320 \times 320 \sim 10^5$ & 3 & 6 & 9 & 11 & 14\\
\hline
$1000 \times 1000 \sim 10^6$ & 3 & 6 & 9 & 11 & 14\\
\hline
\end{tabular}
\caption{Number of updates required to reach the value of $b$ in a evolutionary network for a set of randomly distributed data.}\label{tab:tab1}
\end{table}

Let’s now change the computation order. Instead of updating the system as the network evolves, the evolution takes first place and then, for a fixed network successive updates are done. The results for this case can be seen in Table \ref{tab:tab2}. Here the configurations that are considered are those reached with the evolutionary network when (i) $b <10^{-1}$ and (ii) $b <10^{-5}$.  In these cases, the differences with the evolutionary network are not very significant, but for larger values of the limit of $b$ convergence is slightly faster, and for lower values of this limit convergence is slightly slower in the case (i) and slightly faster in the case (ii). Again, the number of updates slowly increases linearly when $b$ is diminished in both cases (i) and (ii) (Figure \ref{fig:results1}B).

\begin{table}
\begin{tabular}{|c|c|c|c|c|c|}
\hline
\multicolumn{6}{|c|}{Fixed networks}\\
\multicolumn{6}{|c|}{Configuration of evolutionary network  when it reaches: (i)  $b <10^{-1}$, (ii) $b <10^{-5}$.}\\
\hline
$m \times n$ & $b<10^{-1}$ & $b<10^{-2}$ & $b<10^{-3}$ & $b<10^{-4}$ & $b<10^{-5}$ \\
\hline
$32 \times 32 (i) $ & 2 & 5 & 8 & 12 & 16\\
\hline
$320 \times 320 (i)$ & 2 & 5 & 9 & 13 & 17\\
\hline
$32 \times 32 (ii)$ & 2 & 4 & 7 & 9 & 12\\
\hline
$320 \times 320 (ii)$ & 2 & 5 & 7 & 10 & 12\\
\hline
\end{tabular}
\caption{Number of updates required to reach the value of $b$ in two fixed network for a set of randomly distributed data.}\label{tab:tab2}
\end{table}

Table \ref{tab:tab3} shows the results obtained for the same random distribution of data using the fixed network forming a square lattice (cellular automaton with Moore's neighborhood). As can be seen, the number of updates required to achieve a given accuracy rapidly grows with size. To be concrete, the number of updates increases exponentially when $b$ is diminishes (Figure \ref{fig:results1}C) and the size of the network increases (Figure \ref{fig:results1}D).

\begin{table}
\begin{tabular}{|c|c|c|c|c|c|}
\hline
\multicolumn{6}{|c|}{Cellular automaton on square lattice with Moore's neighborhood}\\
\hline
$m \times n$ & $b<10^{-1}$ & $b<10^{-2}$ & $b<10^{-3}$ & $b<10^{-4}$ & $b<10^{-5}$ \\
\hline
$32 \times 32 \sim 10^3$ & 2 & 5 & 8 & 12 & 16\\
\hline
$100 \times 100 \sim10^4$ & 4 & 206 & 1608 & 3350 & 5255\\
\hline
$320 \times 320 \sim 10^5$ & 4 & 419 & 10678 & 29568 & (*)\\
\hline
$1000 \times 1000 \sim 10^6$ & 4 & 420 & 40255 & 109068 & (*)\\
\hline
\end{tabular}
\caption{Number of updates required to reach the value of $b$ by applying a cellular automaton with Moore's neighborhood for a set of randomized data. (*) The stop criterion is reached before obtaining the value of $b$ indicated.}\label{tab:tab3}
\end{table}

\begin{figure}
\begin{center}
  \includegraphics[width=6cm]{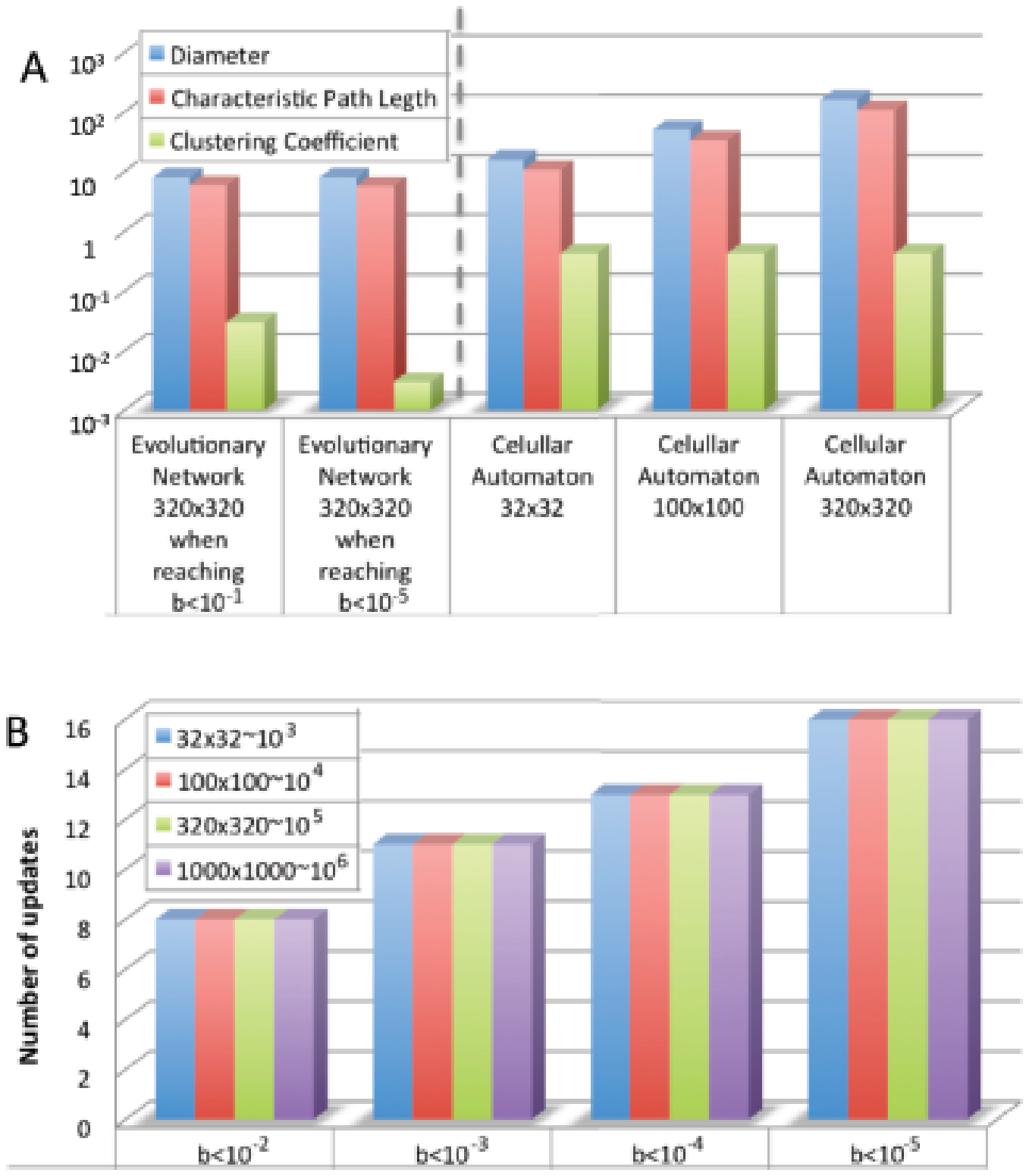}
  \end{center}
  \caption{(A) Topological parameters measured in the analyzed networks. The bar graph shows a constant behavior for the diameter and the characteristic path length in the evolutionary networks (at the left side of the vertical discontinuous line). In the cellular automata case (right side of the vertical discontinuous line), these parameters highly increase with the network size. The clustering coefficient has a low value in the evolutionary network and higher in the cellular automata. Though the connectivity remains always stationary, the evolutionary networks present a much more sparse topology than in the cellular automata cases, which is indicative for a more efficient connectivity in terms of information dispersion. (B) Number of updates required to reach the $b$ value in a specific case using evolutionary networks when the data is not randomly distributed but strongly grouped. In this case the scalability is still very good as in the case with randomly distributed data. The number of updates also slowly increases in a linear way as in the rest of studied cases for the evolutionary network.}
  \label{fig:topol_param}
\end{figure}

To analyze the relationship of these results with the network structure, we have made an analysis of some of the properties of these networks. We have considered the clustering coefficient, the diameter and the characteristic path length (CPL). The clustering coefficient of a node is the ratio $p / r$, where $p$ is the number of links between the neighboring nodes and $r$ is the maximum number of links that would be possible among them; the clustering coefficient of the network is the average value of clustering coefficients of all the network nodes. The network diameter is the maximum distance between two nodes. The characteristic path length gives the expected distance between any two nodes. It is defined as the average number of steps along the shortest paths for all possible pairs of network nodes. As can be seen, the shorter the CPL the better the communication along the network. By construction the number of links of each node in these networks is the same (eight inputs and eight outputs). The clustering coefficient of the evolutionary network has a very low value and that of the cellular automaton is high. The two significant parameters are the diameter of the network and the characteristic path length. The values of these parameters for the fixed networks considered are shown in Table \ref{tab:tab4}. As can be seen, whereas in the four evolutionary network configurations considered the diameter and the characteristic path length have small values, and do not change significantly with the limit of $b$, in the cellular automaton these values are much higher and rapidly growing with size (Figure \ref{fig:topol_param}A). For the cellular automaton convergence is very slow, the number of updates required to reach the limits of $b$ is very high and the scalability is very poor. Therefore this fact suggests that the speed of convergence and scalability of the calculus in the network is strongly associated with the values of these two parameters. This can be seen as manifesting the property of "small world", which appears in different types of complex networks \cite{Albert:2002,Boccaletti:2006}. Topologically speaking, the evolutionary networks are distributed in a more sparse way than the cellular automaton, according to the clustering coefficient values. Regarding CPL and diameter values, the evolutionary networks are much ‘better’ connected than the cellular automaton, in terms of information dissemination. A network with low CPL means that the information contained in one node is more accessible to the rest of the nodes than in a network with high CPL. All this gives a method that re-distributes a network in a way where information is easy to share, and so, computationally speaking, with a low cost in collective computations.

\begin{table}
\begin{tabular}{|p{3cm}|c|p{2.5cm}|c|c|c|}
\hline
\multicolumn{4}{|c|}{Topological Parameters. }\\
\hline
Network & Diameter & Characteristic path length & Clustering coefficient  \\
\hline
Evolutionary network $32 \times 32 $ when reaching $b<10^{-3}$ & 5 & 3.5 & 0.007\\
\hline
Evolutionary network $100 \times 100 $ when reaching $b<10^{-3}$ & 6 & 4.6 & 0.001\\
\hline
Evolutionary network $320 \times 320 $ when reaching $b<10^{-1}$ & 8 & 5.9 & 0.036\\
\hline
Evolutionary network $320 \times 320 $ when reaching $b<10^{-5}$  & 8 & 5.7 & 0.000\\
\hline
Cellular automaton $32 \times 32 $  & 16 & 10.7 & 0.429\\
\hline
Cellular automaton $100 \times 100 $  & 50 & 33.3 & 0.429\\
\hline
Cellular automaton $320 \times 320 $  & 160 & 106.7 & 0.429\\
\hline
\end{tabular}
\caption{Values of some characteristic parameters of the listed networks.}\label{tab:tab4}
\end{table}

\begin{figure}
\begin{center}
  \includegraphics[width=6cm]{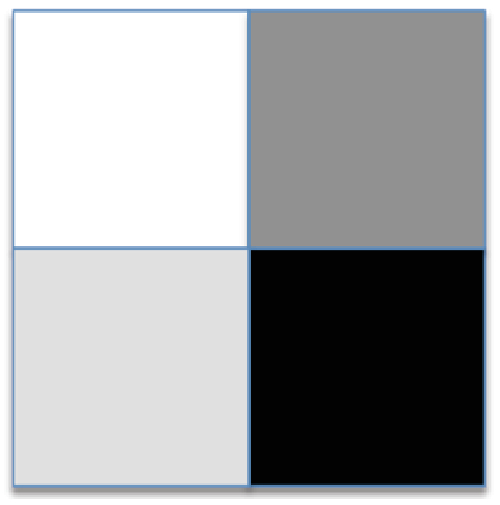}
  \end{center}
  \caption{Schema of data grouped by quadrants.}
  \label{fig:grouped_values}
\end{figure}

Next we have examined the influence of the way in which data are distributed. Instead of a random distribution, we have considered other one in which data is strongly grouped. This consists of four different values distributed in the nodes associated to each of the four quadrants shown in Figure \ref{fig:grouped_values}. The results for the evolutionary network are shown in Table \ref{tab:tab5}. As can be seen scalability is very good as in the case of the data with random distribution, although the number of updates required is somewhat greater than in the case of random data distribution. As in the rest of the studied cases, the number of updates slowly increases in a linear way (Figure \ref{fig:topol_param}B).

\begin{table}
\begin{tabular}{|c|c|c|c|c|}
\hline
\multicolumn{5}{|c|}{Evolutionary network}\\
\hline
$m \times n$ & $b<10^{-2}$ & $b<10^{-3}$ & $b<10^{-4}$ & $b<10^{-5}$ \\
\hline
$32 \times 32 \sim 10^3$ & 8 & 11 & 13 & 16\\
\hline
$100 \times 100 \sim10^4$ & 8 & 11 & 13 & 16\\
\hline
$320 \times 320 \sim 10^5$ & 8 & 11 & 13 & 16\\
\hline
$1000 \times 1000 \sim 10^6$ & 8 & 11 & 13 & 16\\
\hline
\end{tabular}
\caption{Number of updates for an evolutionary network with a set of tightly grouped data.}\label{tab:tab5}
\end{table}

Finally, note that if during the process you add new nodes are added or deleted, the update could proceed as follows:

\begin{itemize}
\item If a node is deleted, its input links could be redirected to its output target nodes, so that those nodes that had access to it now are addressed to those to which the deleted node accessed.
\item If a node is inserted, it provides access to eight of the existing nodes (as input links), and in exchange, each of these nodes will deliver to the new node a link to one of its neighbors (as output nodes).
\end{itemize}

This will allow to perform a system robustness analysis. If the variation in the number of nodes is not very sudden, it is expected the overall system to be not severely impacted, given the information transmission speed. Even in the case of a change in many nodes, it can be expected that a good accuracy in the result of the global parameter will be reached again in a reduced number of updates, i.e. resilience is very good.

\section{Conclusions}
In a networked system, which nodes have a limited access to information only from a few neighbors, the collective computation involving the whole data set is very efficient when the network changes dynamically, or with a fixed structure generated by the same dynamic process. In this case, a fast convergence towards the average value is achieved. Also the number of computing updates shows a good scalability with the size of the network. This is associated with the diameter and the characteristic path length of the network. Given the high rate with which the system converges to the desired value, the system is also is expected to be robust to changes in the number of nodes or in the values distribution assigned to the network. These results clearly contrast with what happens in a fixed regular network in which the convergence is slow and the computation required hugely grows when the size of the system increases.

\section*{Acknowledgement}

This work is partially financed by the Project FIS2008-04120 of the Spanish Ministry for Science and Innovation (MICINN).

\end{document}